\documentclass[prl,aps,floats,twocolumn,showpacs,nofootinbib]{revtex4}

\usepackage{amssymb}
\usepackage{amsmath}

\usepackage{graphicx}
\usepackage{graphics}
\usepackage{dcolumn}
\usepackage{color}
\usepackage{rotate}
\usepackage{fancyhdr} 
\usepackage{hyperref}

\usepackage{umoline}

\def\be{\begin{equation}}
\def\ee{\end{equation}}
\def\bea{\begin{eqnarray}}
\def\eea{\end{eqnarray}}
\def\ba{\begin{eqnarray}}
\def\ea{\end{eqnarray}}
\def\nn{\nonumber}

\def\rmE{\mathrm{E}}

\def\VEV#1{\left\langle #1 \right\rangle}

\newcommand{\hatn}{{\mathbf{\hat n}}}

\newcommand{\bfx}{\mathbf{x}}
\newcommand{\bfk}{\mathbf{k}}
\newcommand{\bfd}{\mathbf{d}}

\definecolor{darkred}{rgb}{.743,0,0}

\newcommand{\refeq}[1]{Eq.~(\ref{eq:#1})}

\begin{document}

\title{Tensor-induced B modes with no temperature fluctuations}

\author{Marc Kamionkowski$^1$, Liang Dai$^1$ and Donghui Jeong$^{2,3}$}
\affiliation{
$^1$Department of Physics and Astronomy, Johns
     Hopkins University, 3400 N.\ Charles St., Baltimore, MD 21218
\\
$^2$Department of Astronomy and Astrophysics, 
The Pennsylvania State University, University Park, PA 16802
\\
$^3$Institute for Gravitation and the Cosmos, The Pennsylvania State University, University Park, PA 16802
}

\date{\today}

\begin{abstract}

The recent indications for a tensor-to-scalar ratio $r\simeq0.2$
from BICEP2 measurements of the cosmic microwave background
(CMB) B-mode polarization present some tension with upper limits
$r\lesssim 0.1$ from measurements of CMB temperature
fluctuations.  Here we point out that tensor
perturbations can induce B modes in the CMB polarization without
inducing {\it any} temperature fluctuations nor E-mode
polarization whatsoever, but only, at the expense of violating
the Copernican principle.  We present this mathematical
possibility as a new ingredient for the model-builder's toolkit
in case the tension between B modes and temperature fluctuations
cannot be resolved with more conventional ideas.
\end{abstract}

\maketitle


The BICEP2 experiment has received considerable attention for
their claimed detection \cite{Ade:2014xna} of a large-angle curl
component (B modes) \cite{Kamionkowski:1996ks,zs} in the
cosmic microwave background (CMB) polarization.  The
interpretation is that these B modes are due to gravitational
waves produced during the time of inflation
\cite{inflation,Rubakov:1982df}.  However, the tensor-to-scalar ratio
$r\simeq0.2$ indicated by BICEP2 CMB polarization is in $\sim2\sigma$
conflict with upper bounds $r\lesssim0.1$ from measurements of
CMB temperature fluctuations \cite{Story:2012wx,Ade:2013zuv}.
The CMB upper limits can be relieved if the scalar spectral
index is allowed to run \cite{Ade:2013uln}, but the running
required is unusually high for single-field slow-roll models.
It may be worthwhile to explore explanations for this
disagreement between the two values of $r$ inferred in case the
tension continues after the dust of the initial detection has
settled.

Here we show that it is possible for a tensor fluctuation to
produce a B mode in the CMB polarization without producing {\it
any} temperature fluctuation nor E-mode polarization whatsoever.
The argument is relatively straightforward given the
total-angular-momentum (TAM) formalism we have developed in
recent work \cite{Dai:2012bc}.  The only catch is that, as we
show, the tensor fluctuation required to do this violates the
Copernican principe; i.e., it implies that we occupy a  preferred
location in the Universe.

We begin by reviewing how a polarization map is decomposed into
E and B modes.  The linear polarization in any direction $\hatn$
on the sky is represented by Stokes parameters $Q(\hatn)$ and
$U(\hatn)$.  However, these quantities are components of a
symmetric trace-free (STF) $2\times2$ tensor, or equivalently, spin-2
field, that lives on the surface of the sky, the two-sphere.  The
polarization can therefore be expanded in terms of
tensor spherical harmonics, a complete set of basis functions
for STF $2\times2$ tensors on the two-sphere.  The linear
polarization is specified at any point by the two Stokes
parameters, and so two sets, E
(for the curl-free part of the polarization field) and B (for
the curl component) are required to provide a complete basis.
The E and B tensor spherical harmonics can be written
\cite{Kamionkowski:1996ks}
\begin{align}
     Y_{(\ell m)AB}^{\rm E}(\hatn) =&
     \sqrt{\frac{2(\ell -2)!}{(\ell +2)!}}
     \left(-\nabla_{A}\nabla_{B}+\frac{1}{2}g_{AB}\nabla^{C}\nabla_{C}\right)
     Y_{(\ell m)}(\hatn) \nn\\
      \equiv& -\sqrt{\frac{(\ell -2)!}{2 (\ell +2)!}}
     W_{AB}^{\rm E} Y_{(\ell m)}(\hatn),   \nn\\ 
     Y_{(\ell m)AB}^{\rm B}(\hatn) =&
     \sqrt{\frac{(\ell -2)!}{2(\ell +2)!}}
     \left(\epsilon_{B}{}^{C}\nabla_{C}\nabla_{A}+\epsilon_{A}{}^{C}
     \nabla_{C}\nabla_{B}\right)Y_{(\ell m)}(\hatn) \nn \\
     \equiv&
     -\sqrt{\frac{(\ell -2)!}{2 (\ell +2)!}} W_{AB}^{\rm B} Y_{(\ell m)}(\hatn),
\label{eqn:EBtensor}
\end{align}
where here $\{A,B\}=\{\theta,\phi\}$, and $\nabla_A$ is a
covariant derivative on the two-sphere, with metric $g_{AB} =
{\rm diag}(1,\sin^2\theta)$ and antisymmetric tensor
$\epsilon_{AB}$.  The last equality
 in each line defines the two
orthogonal tensor-valued derivative operators $W^{\rm E}_{AB}$
and $W^{\rm B}_{AB}$ 
(written down in another form in Eq.~(90) of Ref.~\cite{Dai:2012bc})
 that when applied to the usual scalar
spherical harmonics $Y_{(\ell m)}(\hatn)$ give rise to the E and B
tensor spherical harmonics.
Thus, the two sets of tensor spherical harmonics can be obtained
by applying two orthogonal STF $2\times2$ derivative operators
to the scalar spherical harmonics.  The presence of the
antisymmetric tensor in the definition $Y_{(\ell m)AB}^{\rm B}(\hatn)$
indicates that the B mode and E mode have opposite parities for
fixed $\ell$ and $m$.  It is important to note that the derivative
operators $W^{\rm E}_{AB}$ and $W^{\rm B}_{AB}$ commute with the
total-angular-momentum operator and its third component.
The tensor spherical harmonics are thus eigenstates of the total angular
momentum with quantum numbers $\ell$ and $m$.

We will now develop an analogous decomposition of {\it
three}-dimensional transverse-traceless tensor fields, following
the treatment of Ref.~\cite{Dai:2012bc}.  Before doing so,
however, we review the standard treatment, in which the
transverse-traceless tensor field is decomposed,
\begin{equation}
     h_{ab}(\bfx) = \sum_{\bfk,s} h_{s}(\bfk)
     \widehat{\varepsilon}^s_{ab}(\bfk)e^{i\bfk\cdot\bfx},
\end{equation}
into Fourier components $h_s(\bfk)$ of wavevector
$\bfk$ and polarization $s$ which can be $+$ or $\times$ (and
$\widehat{\varepsilon}^s_{ab}(\bfk)$ are polarization tensors. 
Here $a,b,c,\ldots$ are {\it three}-dimensional spatial indices 
unlike {\it two}-dimensional indices $A,B,C,\cdots$.
This decomposition into Fourier modes can be done because the
plane waves $\widehat{\varepsilon}^s_{ab}(\bfk)e^{i\bfk\cdot\bfx}$ 
constitute a complete othonormal
set of basis functions for a transverse-traceless tensor field.
Power spectra $P_h(k)$ for these tensor perturbations (or
gravitational waves) are given by,
\begin{equation}
     \VEV{h_s(\bfk) h^*_{s'}(\bfk')} = \delta_{ss^{\prime}} (2\pi)^3
     \delta_D(\bfk-\bfk') \frac{P_h(k)}{4},
\label{def:pk_tensor}
\end{equation}

In Ref.~\cite{Dai:2012bc} we showed that transverse-traceless tensor
fields can alternatively be expanded in
transverse-tracless-tensor TAM waves.  To see how this works, we
begin by noting that a scalar function $\phi(\bfx)$ in three
dimensions can be expanded,
\begin{equation}
     \phi(\bfx) = \int\, \frac{d^3k}{(2\pi)^3} \tilde \phi(\bfk)
     \Psi^{\bfk}(\bfx),
\end{equation}
where
\begin{equation}
     \tilde\phi(\bfk) = \int\, d^3x \phi(\bfx) \left[
     \Psi^{\bfk}(\bfx) \right]^*.
\label{def:phixphik}
\end{equation}
in terms of scalar TAM waves
\begin{equation}
     \Psi_{(\ell m)}^k(\bfx) \equiv j_\ell(kr)
     Y_{(\ell m)}(\hatn),
\label{def:Psilmk}
\end{equation}
where $j_\ell(x)$ is the spherical Bessel function of the first kind. These TAM waves
are eigenfunctions of the three-dimensional Laplacian $\nabla^2$
with eigenvalue $-k^2$ and also eigenfunctions of the total
angular momentum and its third component with quantum numbers
$\ell$ and $m$ respectively.  These scalar TAM waves constitute a
complete orthonormal basis for scalar functions. The orthonormality
 relation for the basis functions is
\begin{equation}
     \int\ d^3x\, \left[\Psi_{(\ell m)}^k(\bfx)\right]^*
     \Psi_{(\ell' m')}^{k'}(\bfx)
     =\frac{\delta_{\ell \ell'}\delta_{mm'}}{16\pi^2} \frac{(2\pi)^3}{k^2}
     \delta_D(k-k'),
\end{equation}
where $\delta_{ij}$ is the Kronecker delta.  Completeness is
demonstrated by
\begin{equation}
     \sum_{\ell m} \int \frac{k^2\, dk}{(2\pi)^3} \left [4 \pi i^\ell
     \Psi^k_{(\ell m)} (\bfx) \right]^* \left [4 \pi i^\ell
     \Psi^k_{(\ell m)} (\bfx') \right] = \delta_D(\bfx-\bfx').
\end{equation}

Just as the E/B tensor spherical harmonics were obtained by
applying appropriately defined tensor-valued derivative
operators, transverse-traceless-tensor TAM waves can be obtained
by applying transverse-traceless-tensor--valued derivative
operators to the scalar TAM waves.  The appropriate derivative
operators are
\begin{eqnarray}
     T_{ab}^{\rm B}  &=&
     K_{(a}M_{b)}+M_{(a}K_{b)}+2D_{(a}K_{b)},  \nonumber \\
     T_{ab}^{\rm E}  &=& M_{(a}M_{b)}-K_{(a}K_{b)}+2D_{(a}M_{b)},
\end{eqnarray}
where we define operators,
\be
D_a\equiv\frac{i}{k}\nabla_a,\quad K_a\equiv -iL_a,\quad
M_a\equiv\epsilon_{abc}D^b K^c,
\label{eqn:operatordefs}
\ee
 for the subspace of $\nabla^2=-k^2$, and $L_a=-i \epsilon_{abc}
 x^b \nabla^c$ is the orbital-angular-momentum operator. 
The two tensor operators
$T_{ab}^{\rm B}$ and $T_{ab}^{\rm E}$ are orthogonal.  They are
also both transverse and traceless, and they both commute with
the Laplacian and with the total angular momentum and its
third component.  Thus, the transverse-traceless-tensor TAM
waves,
\begin{eqnarray}
\label{eq:spin-two-EB-base-TAM}
     \Psi_{\left(\ell m\right)ab}^{{\rm B},k}(\bfx)&=& -
     \sqrt{\frac{(\ell -2)!}{2 (\ell +2)!}} 
     T_{ab}^{{\rm} B}\Psi^k_{\left(\ell m\right)}(\bfx),\nn\\
     \Psi_{\left(\ell m\right)ab}^{{\rm E},k}(\bfx)&=&
     -\sqrt{\frac{(\ell -2)!}{2 (\ell +2)!}} 
     T_{ab}^{{\rm} E}\Psi^k_{\left(\ell m\right)}(\bfx),
\end{eqnarray}
constitute a complete basis for transverse-traceless tensor
fields in three dimensions.  Note that $T_{ab}^{\rm E}$ is even
under a parity inversion, while $T_{ab}^{\rm B}$ is odd.  The E
and B tensor TAM waves thus have opposite parity for the same
$\ell$ and $m$.

Thus, the most general three-dimensional transverse-traceless
tensor field can be written,
\begin{eqnarray}
     h_{ab}(\bfx) &=& \sum_{\ell m}\int \,\frac{k^2\, dk}{(2\pi)^3} 4
     \pi i^\ell  \left[ h_{\ell m}^{\rm E}(k) \Psi^{{\rm E},k}_{(\ell m)ab}(\bfx) \right.
     \nn\\ 
     &  & \left. +
      h_{\ell m}^{\rm B}(k) \Psi^{{\rm B},k}_{(\ell m)ab}(\bfx) \right].  
\end{eqnarray}

An alternative set of basis TAM waves can be derived in which
the TAM waves have fixed helicity.  These helicity $s=\pm2$ TAM
waves will be useful below and are related to the E/B TAM waves
through 
$\Psi^{s=\pm 2,k}_{(\ell m)ab}(\bfx)=[\Psi^{{\rmE},k}_{(\ell m)ab} \pm i \Psi^{{\rm B},k}_{(\ell m)ab}]/\sqrt2$. 
They allow the most general transverse-traceless tensor field to
be expanded,
\bea
h_{ab}(\bfx) = \sum_{\ell m} \sum_{s=\pm2} \int \,\frac{k^2\, dk}{(2\pi)^3} 4
     \pi i^\ell h^s_{\ell m}(k) \Psi^{s,k}_{(\ell m)ab}(\bfx).
\eea

In Ref.~\cite{Dai:2012bc} it was proved for scalar fluctuations
that scalar TAM waves with quantum numbers $\ell$ $m$ contribute
only to CMB (and
any other observable) power spectra of the same $\ell$.  This is a
consequence of the fact that the TAM waves and the
scalar/vector/tensor spherical harmonics transform as
representations of the rotation group of order $\ell$.  The same
will be true for tensor TAM waves as well:  i.e., tensor TAM
waves of order $\ell$ will contribute only to observable power
spectra of multipole moment $\ell$.  More importantly for our
purposes, though, a similar statement applies to the E/B
decomposition. The argument is based on the parity transformation,
i.e., the dependence of the basis functions under a flip of the
displacement vector $\bfx \rightarrow -\bfx$ with respect to the
location of the observer.
The E-mode tensor TAM wave, scalar spherical
harmonic, and E-mode tensor spherical harmonic all have parity
$(-1)^{\ell}$, while the B-mode tensor TAM wave and B-mode tensor
spherical harmonic transform as $(-1)^{\ell+1}$.  Since the physics
(Thomson scattering and radiative transfer) involved in the
generation of CMB fluctuations is parity conserving, {\it the
B-mode component of the tensor field gives rise to B-mode
polarization but contributes nothing to the temperature
fluctuation nor the E-mode polarization}.
The converse is true for the E-mode component of the tensor field.
As an obvious
corollary, if B-mode TAM-wave components have larger power than the E-mode components do, the observed B-mode polarization power spectrum may be
larger than would be inferred from the temperature and E-mode polarization
power spectra.

As we now demonstrate, an excess of B-mode power over E-mode
power in the TAM-wave expansion requires a violation of the
Copernican principle.  This is because the decomposition of the
tensor field into E and B modes is not translation invariant.
This can be seen from the following physics argument:  CMB
polarization is generated by Thomson scattering of temperature
fluctuations.  B-mode polarization therefore requires that there
be a temperature fluctuation at the surface of last scatter,
even though, as argued above, a tensor B-mode TAM wave produces
no temperature fluctuation for an observer at the origin.  Thus,
the decomposition into E- and B-mode tensor TAM waves is not
translationally invariant.

More explicitly, one can show that a B-mode TAM wave defined
with respect to a given origin has nonzero functional overlap
with an E-mode TAM wave defined with respect to a different
origin. To avoid technical complications, 
here we demonstrate for spin-one, transverse vector field; 
the statement is also true for a transverse-traceless tensor field. 
The vector analogue of \refeq{spin-two-EB-base-TAM} is simply
\bea
\Psi^{{\rm B},k}_{(\ell m)a}(\bfx) & = &  \left[\ell(\ell+1)\right]^{-1/2} K_a(\bfx) \Psi^k_{(\ell m)}(\bfx), \nn\\
\Psi^{{\rm E},k}_{(\ell m)a}(\bfx) & = & \left[\ell(\ell+1)\right]^{-1/2} M_a(\bfx) \Psi^k_{(\ell m)}(\bfx),
\eea
where the vector operators $K_a$ and $M_a$ explicitly depend
on the choice of coordinate origin. We can compute the overlap
between E- and B-mode TAM waves defined with respect
to two different coordinate origins separated by a constant vector $\bfd$.
Taking identical $k$, $\ell$ and $m$, for example,
\bea
&& \int d^3 x \, \left[ \Psi^{{\rm B},k}_{(\ell m)a}(\bfx+\bfd) \right]^*   \Psi^{{\rm E},k,a}_{(\ell m)}(\bfx) \nn\\
& \propto & \int d^3 x \, \left[ K_a(\bfx+\bfd) \Psi^k_{(\ell m)}(\bfx+\bfd) \right]^*   M^a(\bfx) \Psi^k_{(\ell m)}(\bfx) \nn\\
& = & - \int d^3 x \, \left[\Psi^k_{(\ell m)}(\bfx+\bfd)\right]^* \left[ K_a(\bfx+\bfd)  M^a(\bfx) \Psi^k_{(\ell m)}(\bfx) \right]. \nn\\
\eea
Now the two operators act successively on the scalar TAM
wave because of the anti-Hermitian condition 
$[K_a(\bfx)]^\dagger=-K_a(\bfx)$. Note that $K_a(\bfx+\bfd)M^a(\bfx)$
does not vanish if $\bfd\neq0$, while $K_a(\bfx)M^a(\bfx)=0$
is true. Simple algebra gives,
\bea
&& K_a(\bfx+\bfd)M^a(\bfx) = \left( K_a(\bfx) - \epsilon_{abc}\, d^b \nabla^c \right) M_a(\bfx),\nn\\
& = & - \epsilon_{abc}\, d^b \nabla^c M_a(\bfx) = i k\, d^b K_b(\bfx).
\eea
If $\bfd$ is along the $z$-direction so that $d^a K_a = - i d L_z$, we have
\bea
&& \int d^3 x \, \left[ \Psi^{{\rm B},k}_{(\ell m)a}(\bfx+\bfd) \right]^*   \Psi^{{\rm E},k,a}_{(\ell m)}(\bfx) \nn\\
& \propto &  - \int d^3 x \left[\Psi^k_{(\ell m)}(\bfx+\bfd)\right]^* k\, d L_z \Psi^k_{(\ell m)}(\bfx) \nn\\
& = & - m k\,d \int d^3 x \left[\Psi^k_{(\ell m)}(\bfx+\bfd)\right]^* \Psi^k_{(\ell m)}(\bfx).
\eea
The overlap between two scalar TAM waves of shifted coordinate
origins is in general nonzero. This implies that even if the 
power of E-mode TAM waves might vanish when the E/B-decomposition
is performed with respect to one observer, the decomposition 
yields nonzero E-mode power with respect to another observer at 
a different location in the Universe. 

We now show in a different way that statistical homogeneity
requires the power in E and B TAM transverse-traceless-tensor
TAM waves to waves to be the same.  We first relate the power
spectra of TAM-wave coefficients to those of Fourier
coefficients.  Plane-wave tensor states of polarization $+$ and
$\times$ can be added to construct helicity plane waves with
Fourier amplitudes $h_s(\bfk)$ labelled by the wave vector
$\bfk$ and the helicity $s=\pm2$.  If statistical homogeneity
(or translation invariance) is respected, 
then $\VEV{h_s(\bfk) h^*_{s'}(\bfk')} \propto \delta_D(\bfk
-\bfk')$ (it is not required to be proportional to
$\delta_{ss'}$ though). On the other hand, in
terms of the TAM-wave coefficient,
\be
\VEV{h^{\alpha}_{\ell m}(k) [ h^{\beta}_{\ell' m'}(k') ]^*} 
= P_{\alpha\beta}(k)\delta_{\ell \ell'} \delta_{mm'} \frac{(2\pi)^3}{k^2} \delta_D(k-k'),
\ee
with $\alpha,\beta=\{E,B\}$.
We have assumed diagonalization with respect to
angular-momentum quantum numbers to preserve statistical
isotropy. Note that parity invariance would require $P_{\rm
EB}(k)=0$. Transforming to the helcity TAM basis
$h^{s=\pm2}_{\ell m}(k) = [h^{\rm E}_{\ell m}(k)\mp i h^{\rm
B}_{\ell m}(k)]/\sqrt2$, we have
\bea
\VEV{h^s_{\ell m}(k) [ h^{s'}_{\ell' m'}(k') ]^*} & = & P_{ss'}(k) \delta_{\ell \ell'} \delta_{mm'} \frac{(2\pi)^3}{k^2} \delta_D(k-k'), \nn\\
\eea
where the $2 \times 2$ matrix is given by
\bea
P_{ss'}(k) = \left(\begin{array}{cc}
P_{22}(k) & P_{2,-2}(k)\\
P_{-2,2}(k) & P_{-2,-2}(k)
\end{array}\right),
\eea
with
\bea
P_{22}(k) & = & \frac12 \left[ P_{\rm EE}(k) + P_{\rm BB}(k) \right] - \mathfrak{Im} P_{\rm EB}(k), \nn\\
P_{-2,-2}(k) & = & \frac12 \left[ P_{\rm EE}(k) + P_{\rm BB}(k) \right] + \mathfrak{Im} P_{\rm EB}(k), \nn\\
P_{2,-2}(k) & = & [ P_{-2,2}(k) ]^* \nn\\
& = & \frac12 \left[ P_{\rm EE}(k) - P_{\rm BB}(k) \right] \mp i \mathfrak{Re} P_{\rm EB}(k).
\eea
Now using the relation,
\bea
h_s(\bfk) = \sum_{\ell m}\,\left[{}_{-s}Y_{(\ell
m)}(\hat\bfk)\right]^* h^s_{\ell m}(k),
\eea
we find the power spectra for Fourier components,
\bea
&& \VEV{h_s(\bfk) [h_{s'}(\bfk')]^*} = P_{ss'}(k) \frac{(2\pi)^3}{k^2} \delta_D(k-k') \nn\\
&& \times \sum_{\ell m}\,\left[{}_{-s}Y_{(\ell m)}(\hat\bfk') \right]^* {}_{-s'}Y_{(\ell m)}(\hat\bfk).
\eea
Only for $s=s'$ does the summation over $\ell,m$ add up to
$\delta_D(\hat\bfk - \hat\bfk')$, so that the power spectrum
$\VEV{h_s(\bfk) [h_{s'}(\bfk')]^*} \propto \delta_D(k-k')
\delta_D(\hat\bfk - \hat\bfk')/k^2 = \delta_D(\bfk - \bfk')$. We
therefore conclude that the Copernican principle requires
$P_{2,-2}(k)=0$; i.e. $P_{\rm EE}(k)=P_{\rm BB}(k)$ and the real
part of $P_{\rm EB}(k)$ vanishes, {\it whether or not} the
two-point statistics are parity invariant.

To conclude, the $\sim2\sigma$ tension between the
BICEP2 value for $r$ and CMB upper limits is probably not
sufficiently significant to warrant abandoning the Copernican
principle.  Still, the possibility discussed here to produce a
larger B-mode signal, without inducing any temperature
fluctuations or E-mode polarization, may prove to be useful
should the $\sim2\sigma$ disagreement continue to be established
with greater signficance.  And perhaps this may be useful in an
explanation that does not involve a violation of the Copernican
principle.  For example, perhaps a theory in which the
amplitudes of the E- and B-mode tensor TAM waves have a highly
non-Gaussian distribution.  There may then be an {\it apparent}
violation of the Copernican principle in our observable Universe
that renders the B-mode TAM-wave amplitude larger, relative to
the E-mode TAM-wave amplitude, than would be expected with a
Gaussian distribution.  We leave the construction of a model to
implement this idea to future work.

\smallskip
This work was supported by NSF Grant No.\ 0244990 and by the John
Templeton Foundation.

\end{document}